\documentclass[prb,twocolumn,showpacs,citeautoscript,amsmath,amssymb]{revtex4}

\usepackage{graphicx}
\usepackage{dcolumn}
\usepackage{bm}

\begin{document}


\title{Controlling the density of the 2DEG at the SrTiO$_3$/LaAlO$_3$ interface}

\author{A. Janotti}
\author{L. Bjaalie}
\author{L. Gordon}
\author{C. G. Van de Walle}
\affiliation{Materials Department, University of California, Santa Barbara, CA 93106-5050}

\date{\today}

\begin{abstract}

The polar discontinuity at the SrTiO$_3$/LaAlO$_3$ interface (STO/LAO) can in principle sustain an electron density of 3.3$\times$10$^{14}$ cm$^{-2}$ (0.5 electrons per unit cell).  However, experimentally observed densities are more than an order of magnitude lower.  Using a combination of first-principles and Schr{\" o}dinger-Poisson simulations we show that the problem lies
in the asymmetric nature of the structure, i.e., the inability to form a second LAO/STO interface that is a mirror image of the first, or to fully passivate the LAO surface.
Our insights apply to oxide interfaces in general, explaining for instance why the SrTiO$_3$/GdTiO$_3$ interface has been found to exhibit the full density of 3.3$\times$10$^{14}$ cm$^{-2}$.
\end{abstract}


\maketitle

The realization of a two-dimensional electron gas (2DEG) at the SrTiO$_3$/LaAlO$_3$ interface (STO/LAO) has set off an explosion of interest in oxide electronics.
This 2DEG exhibits densities that are difficult to achieve in conventional semiconductors \cite{Ohtomo04,Mannhart08,Huijben09} and displays unique behavior including ferromagnetism \cite{Brinkman07}, superconductivity \cite{Reyren07}, and even the puzzling coexistence of both \cite{Li11}. It has been proposed as the basis for novel electronic devices that exploit strong electron-electron correlation in the narrow bands derived from $d$ states of the transition metal \cite{Hwang12}.
STO/LAO heterostructures 
have been fabricated using methods that allow unprecedented control over layer thickness, such as pulsed layer deposition and molecular beam epitaxy \cite{Mannhart08,Huijben09}.
While great progress has been made in characterization and exploitation of the physical phenomena, the mechanisms that determine the density of electrons in the 2DEG have remained a subject of intense debate \cite{Park06,Siemons07,Demkov08,Popovic08,Pentcheva10,Bristowe11,Sukit11}. This lack of understanding inhibits achieving the control that is required for device applications.

\begin{figure}[h!]
\includegraphics[width=3.2in]{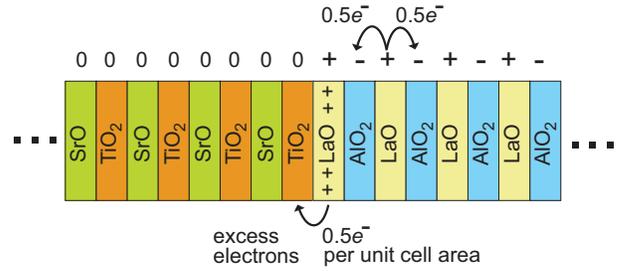}
\caption{\label{fig:layers}
(Color online)
Layer structure of an STO/LAO heterostructure with TiO$_2$-LaO planes at the interface.   Nominal charges are indicated above each layer. LaO planes act as electron donors; the TiO$_2$ plane terminating STO is charge neutral, and therefore the interfacial LaO plane acts as a delta-doped layer of donors with a density of 0.5 electrons per unit cell.}
\end{figure}

Typically, an LAO layer of less than 20 nm is deposited on a TiO$_2$-terminated [001]-oriented STO substrate or epilayer
\cite{Mannhart08,Huijben09}, as shown in Fig.~\ref{fig:layers}.
The electrical conductivity at the buried interface is then probed as a function of temperature \cite{Mannhart08,Huijben09}.  Carrier densities up to 2$\times$$10^{13}$ cm$^{-2}$ have been reported \cite{Siemons07,Basletic08,Kalabukhov07,Thiel06}.   Low-temperature sheet resistance varying from $10^{-2}$ to $10^{4}$ $\Omega/\square$ has been observed, displaying a strong dependence on the oxygen partial pressure in the growth environment or post-growth annealing treatments \cite{Brinkman07,Siemons07,Kalabukhov07,Herranz07}.  A dependence of the sheet resistivity and carrier density on the thickness of the LAO layer has also been observed:  heterostructures with LAO layers less than 4 unit cells ($\sim$1.6 nm) thick exhibit insulating behavior, while thicker layers become conducting \cite{Thiel06}.  However, experiments by Huijben {\em et al.} \cite{Huijben06} indicated the existence of conducting interfaces for LAO layers as thin as two unit cells in the presence of an STO layer on top of the LAO.

A variety of models have been put forth to explain the origin of the carriers at the interface \cite{Park06,Siemons07,Demkov08,Popovic08,Pentcheva10,Bristowe11,Sukit11}.
However, none can account for all the experimental observations, raising questions about their validity and general applicability.
The prevailing models assume that the carriers originate from somewhere other than the ideal interface, such as from the supposedly negatively charged LAO top surface (driven by the ``polar catastrophe'') \cite{Demkov08,Pentcheva10}, from oxygen vacancies \cite{Siemons07,Bristowe11,Sukit11,Herranz07,Kalabukhov07}, or from Sr-La intermixing at the interface \cite{Nakagawa06,Qiao10}.

Here we note that the source of the carriers should not be in question.  As explained by Fig.~\ref{fig:layers}, the polar discontinuity at the STO/LAO interface provides an intrinsic source of electrons with a density of 0.5 electrons ($e^-$) per unit cell (3.3$\times$$10^{14}$ cm$^{-2}$).
In an ionic picture (based on Sr$^{2+}$, La$^{3+}$, Ti$^{4+}$, and O$^{2-}$), LaAlO$_3$ can be regarded as a stacking of (LaO)$^+$ and (AlO$_2$)$^-$ planes along the [001] direction; LaO planes donate electrons which become bound in the neighboring AlO$_2$ planes.  SrTiO$_3$, on the other hand, is composed of alternating charge-neutral (SrO)$^0$ and (TiO$_2$)$^0$ planes.  At the interface with TiO$_2$-terminated STO, the TiO$_2$ layer is already charge-neutral, causing the interfacial LaO layer to act as a sheet of donors, donating 0.5$e^-$ per unit cell, which due to the large conduction-band offset between LAO and STO flow into the STO.
The question is therefore not ``where do the carriers come from?'', but rather, ``where do the electrons disappear to?''  I.e., why is
the observed carrier density an order of magnitude lower \cite{Brinkman07,Siemons07,Basletic08,Kalabukhov07} than the expected 0.5$e^-$ per unit cell (3.3$\times$$10^{14}$ cm$^{-2}$)?

Based on first-principles calculations combined with Schr{\"o}dinger-Poisson (SP) simulations we attribute the problem to the lack of a suitable termination for the top surface of the LAO layer.  The asymmetric nature of the resulting layer structure drains electrons away from the 2DEG, and further exposes the heterostructure to the detrimental effects of point-defect formation.
Our insights enable us to propose specific strategies for overcoming the problems associated with the current STO/LAO structures, and also provide guidelines for the choice of other materials combinations.  We note that 2DEG densities as high as 3.3$\times$10$^{14}$ cm$^{-2}$
have already been observed at SrTiO$_3$/GdTiO$_3$ interfaces, consistent with the predictions of our model.

Our first-principles calculations are based on generalized Kohn-Sham theory within the projector-augmented wave method as implemented in
the VASP code \cite{dft,kressepaw,vasp2}.
We use the hybrid functional of Heyd, Scuseria, and Ernzerhof (HSE) \cite{hse}, which has been shown to produce electronic structure and band gaps in closer agreement with experiment and a better description of charge localization \cite{varley12}, compared to conventional functionals based on the local density approximation (LDA) or the generalized gradient approximation (GGA).
For the SrTiO$_3$ and LaAlO$_3$ cubic perovskite phases we use a 4$\times$4$\times$4 special k-point set for integrations over the Brillouin zone and an energy cutoff of 500 eV for the plane-wave basis set.
For the heterostructures, the calculations were performed for a (STO)$_{8.5}$/(LAO)$_{7.5}$ superlattice containing two equivalent TiO$_2$-LaO interfaces, with a 4$\times$4$\times$1 k-point set and a 400 eV energy cutoff.  Tests using 6$\times$6$\times$1 and 8$\times$8$\times$1 meshes resulted in changes of less than 0.1 eV in Fermi energy, and charge densities remained essentially unchanged. The in-plane lattice constant was fixed to that of STO, representing a heterostructure coherently grown on an STO substrate.  Full relaxation was allowed for both the out-of-plane lattice constant and all atomic positions.

The SP simulations were performed using the {\it nextnano$^3$} simulation software, which
solves for the electrostatic potential, charge density, and Fermi level across the heterostructures.  The input parameters include electron effective masses (1.0 $m_e$, fitted to reproduce the first-principles density of states) and dielectric constants (300 for STO, 27 for LAO).   The valence-band band offset was set to the first-principles calculated value of 0.1 eV (higher in LAO).  Background doping  at a level $n$=10$^{17}$ cm$^{-3}$ in the STO was assumed.  In the SP simulations we do not distinguish between the different types of carriers contributing to the 2DEG density, i.e., if $d_{xy}$ or $d_{yz}$/$d_{xz}$.

\begin{figure}[h!]
\includegraphics[width = 2.6 in]{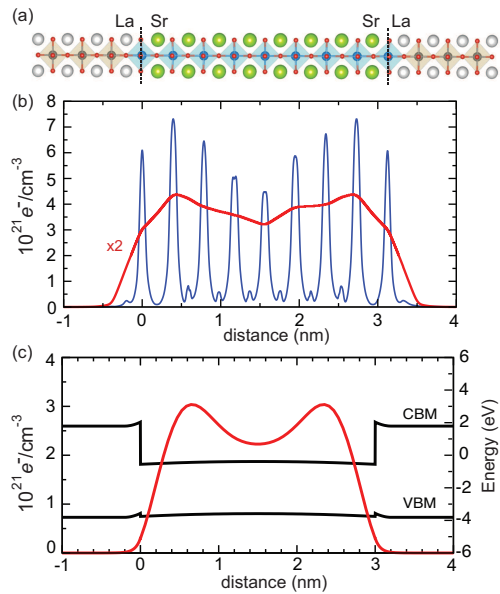}
\caption{\label{fig:STOLAO}
(Color online)
First-principles and Schr\"{o}dinger-Poisson results for a (SrTiO$_3$)$_{8.5}$/(LaAlO$_3$)$_{7.5}$ superlattice with TiO$_2$-LaO interfaces.
(a) Atomic structure, with oxygen atoms shown in red.  Ti-centered octahedra are shown.
(b) First-principles planar and macroscopically averaged charged density of the occupied subbands (excluding the STO and LAO valence bands).
(c) Schr\"{o}dinger-Poisson simulations for the same superlattice, showing good agreement with the first-principles results.  CBM stands for conduction-band minimum and VBM for valence-band maximum.}
\end{figure}

The first-principles calculations quantitatively confirm that the free carriers in the 2DEG at the interface originate from the interfacial LaO plane.
Figure~\ref{fig:STOLAO} shows the integrated charge density, plotted along the [001] direction, associated with the electrons generated at the interface.   This charge density appears exclusively on the STO side and corresponds to occupied subbands in the conduction band.  The integrated charge is 3.3$\times$10$^{-14}$ cm$^{-2}$ per interface (i.e., 0.5$e^-$ per unit cell area), exactly what we expect based on the consideration of the interfacial LaO as a delta-doped donor layer.
We observe that this charge is delocalized over multiple planes of Ti and {\it not} localized on a single Ti layer at the interface.  If the latter were the case \cite{Pentcheva10}, the electrons would be immobilized on interfacial Ti$^{3+}$, which would be inconsistent with the observation of a high-mobility 2DEG.

The symmetric nature of the charge density in the STO layer arises from the fact that our first-principles calculation needs to maintain periodicity along the [001] direction, and therefore corresponds to a superlattice containing two interfaces.  We have verified, however, that the results described here do not depend on the thickness chosen for the STO and LAO layers.

%


Very similar results are produced by SP simulations \cite{nextnano} for the (STO)$_{8.5}$/LAO$_{7.5}$ superlattice.
The nominal charge density of 0.5$e^-$ per unit cell area, as obtained from first-principles calculations, is reproduced in the SP simulation for the (STO)$_{8.5}$/LAO$_{7.5}$ superlattice, as shown in Fig.~\ref{fig:STOLAO}(c).   While these simulations do not include the intricacies of the STO conduction-band structure, they do accurately capture the overall carrier density and distribution near the interface, which is the focus of our study.   The SP simulations allow modeling systems with larger dimensions and, more importantly, lacking periodicity, and hence enable us to study layer structures that are beyond the capabilities of the first-principles calculations.

\begin{figure}
\includegraphics[width = 2.8in]{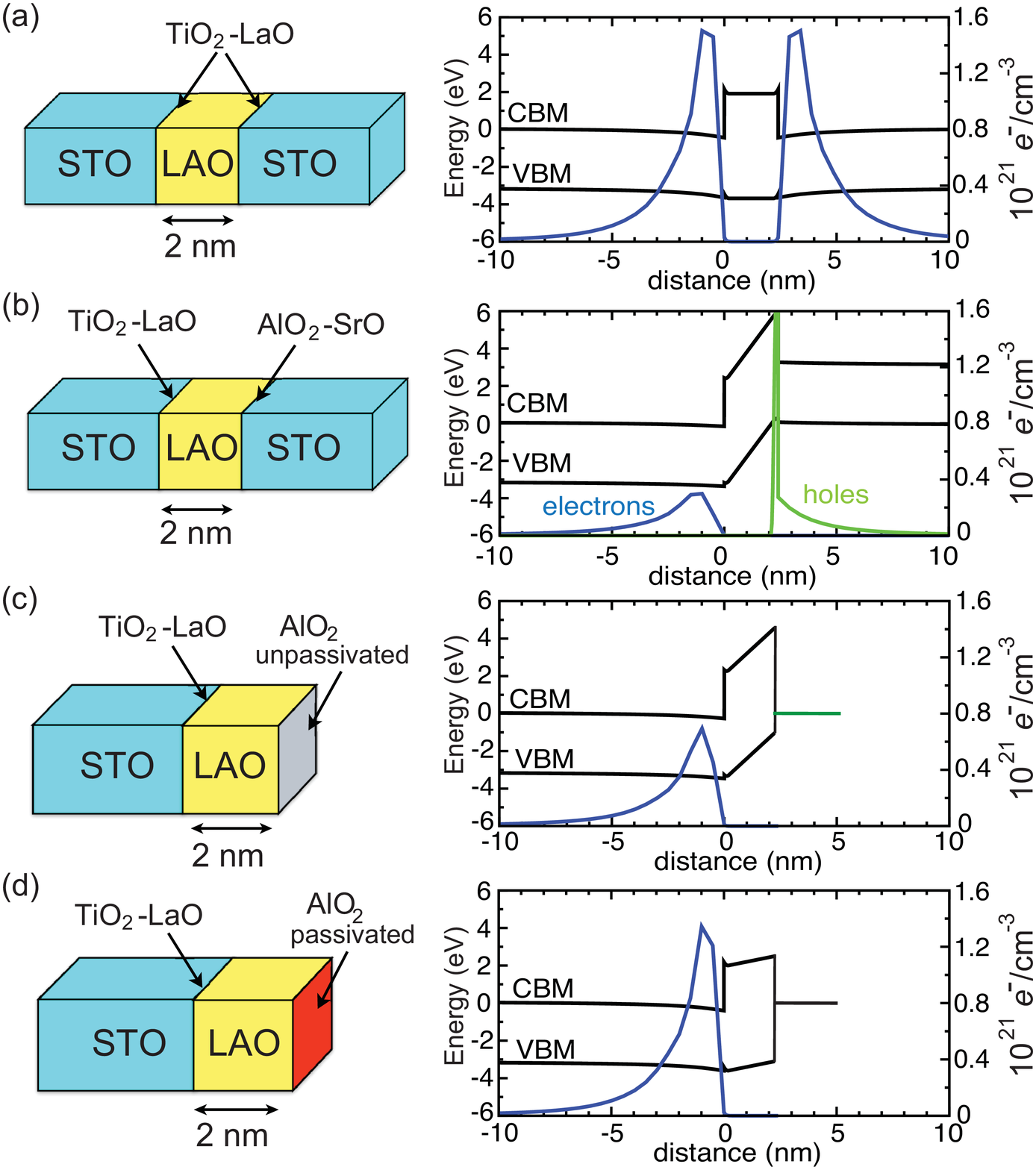}
\caption{\label{fig:SP}
(Color online)
Schr{\"o}dinger-Poisson simulations for SrTiO$_3$/LaAlO$_3$ interfaces. Layer structures are depicted on the left, and the corresponding band diagrams and charge density distributions on the right.  The zero of energy is placed at the Fermi level.
(a) STO/LAO/STO with two equivalent TiO$_2$-LaO interfaces. The integrated electron density is 3.3$\times$10$^{14}$ cm$^{-2}$ per interface.
(b) STO/LAO/STO with inequivalent interfaces: TiO$_2$-LaO on the left, AlO$_2$-SrO on the right.
(c), STO/LAO with TiO$_2$-LaO at the interface and an AlO$_2$-terminated surface containing acceptor-like surface states (green horizontal bar).
(d) STO/LAO with TiO$_2$-LaO at the interface and a passivated surface.
}
\end{figure}

Given that our simulations so far indicate that the 2DEG density should be expected to correspond to 0.5$e^-$ per unit cell, we now address the question of why experiment shows much lower densities \cite{Brinkman07,Siemons07,Basletic08,Kalabukhov07}.   The answer lies in the fact that the type of symmetric structure depicted in Fig.~\ref{fig:STOLAO}, with two identical interfaces, is never achieved experimentally.
In most practical implementations, the LAO layer is of finite thickness and has a surface terminated on an AlO$_2$ plane.  The consequences are examined in Fig.~\ref{fig:SP} for a 2-nm thick LAO layer (about 5 unit cells) on thick STO.  In panel a it is assumed that a perfect STO layer (with TiO$_2$-LaO planes at the interface) can be deposited on top of the LAO. This effectively reproduces the symmetric situation that was investigated in Fig.~\ref{fig:STOLAO}, the only difference being the smaller LAO thickness and larger STO thickness; reassuringly, the results are very similar, with the full density corresponding to 0.5$e^-$ per unit cell appearing in the 2DEG.  Note that the electrostatic potential (reflected in the slope of the conduction band) is essentially flat across the LAO layer, indicating no charge is being transferred between the two interfaces.

The case depicted in Fig.~\ref{fig:SP}(a) is unrealistic, because experimentally it has turned out to be difficult (or even impossible) to grow STO/LAO/STO structures with an interface between LaO and TiO$_2$ planes on the right-hand side \cite{Huijben06}.
Figure~\ref{fig:SP}(b) depicts the situation for an STO/LAO/STO layer structure with an interface between AlO$_2$ and SrO on the right.  By similar logic as applied to the TiO$_2$-LaO interface leading to donor doping,  an AlO$_2$-SrO interface leads to acceptor doping with a sheet density of 3.3$\times$10$^{-14}$ cm$^{-2}$ since the AlO$_2$ layer is lacking 0.5$e^-$ per unit cell which (in the bulk) would come from an LaO plane (the SrO plane in STO being charge-neutral).
The electrons at the TiO$_2$-LaO interface are higher in energy than the holes at the AlO$_2$-SrO interface, making this situation unstable.  The energy of the system is lowered by transferring electrons from the TiO$_2$-LaO interface to the AlO$_2$-SrO interface, leading to a strong dipole across the LAO layer (note the slope in electrostatic potential).
In the process, the 2DEG density is drastically reduced, although not to zero because the electric field in the LAO layer results in the valence band of LAO rising above the Fermi level (with a hole gas appearing at the right-hand interface), thus limiting the transfer of electrons.

Having an ideal AlO$_2$-SrO interface on the right [Fig.~\ref{fig:SP}(b)] is similar to terminating the LAO layer with an ideal AlO$_2$ surface---indeed, all planes in the STO on top are charge-neutral and hence do not contribute to any charge exchange.  The only difference is that a realistic AlO$_2$-terminated LAO surface would exhibit oxygen dangling bonds, giving rise to partially filled surface states with energies in the lower part of the band gap.  Similar charge transfer would occur as described for the case depicted in Fig.~\ref{fig:SP}(b), except that the acceptor states are now {\it deep} acceptor states (with an ionization energy of 1 eV above the LAO VBM), as shown in Fig.~\ref{fig:SP}(c). The Fermi level at the surface will be pinned at these acceptor states, limiting the rise of the LAO VBM.  This is actually beneficial since it limits the amount of electron transfer out of the 2DEG, compared to the situation of Fig.~\ref{fig:SP}(b).  Note that this corresponds to a decrease in the slope of the potential, i.e., the magnitude of the electric field, across the LAO layer: the smaller the field, the higher the 2DEG density.

This suggests a strategy for increasing the electron density in the 2DEG, namely minimizing the slope in the potential across the LAO.
Figure~\ref{fig:SP}(d) depicts a fully passivated surface, i.e., the density of acceptors at the surface is assumed to be zero.
While there is a still a slight slope in the potential (due to the fact that the centers of gravity of the positive and negative charge distributions do not coincide), the 2DEG density now recovers its nominal value of 0.5$e^-$ per unit cell.

Our findings easily explain the experimentally observed dependence of 2DEG density on LAO thickness \cite{Thiel06}.  As noted in the discussion of Fig.~\ref{fig:SP}(b) above, full transfer of electrons to the top surface will tend to occur if the LAO layer thickness is insufficient to bring the VBM at the surface above the Fermi level.  Given a positive sheet charge density of 3.3$\times$10$^{-14}$ cm$^{-2}$ at the STO/LAO interface, Gauss' law predicts a field of about 0.25 V/{\AA}, and thus it takes a ``critical thickness'' of about 3 or 4 unit cells of LAO to develop enough of an increase in potential to bring the VBM (or acceptor-like surface states) of LAO above the Fermi level, at which point transfer of electrons out of the 2DEG is suppressed and observable mobile charge appears in the 2DEG.  This is confirmed by explicit SP simulations as a function of LAO layer thickness (not shown).  
In principle the carrier density in the 2DEG should continue to rise as a function of LAO thickness.  However, growing thicker LAO layers of high quality will be difficult because an increasingly large fraction of the layer has the Fermi level lying close to the VBM, a condition that is conducive to formation of oxygen vacancies.  These will both interfere with the charge balance and degrade the quality of the layer.  All of this is completely consistent with results in the literature; our main point is that complicated arguments (e.g., relating to interfacial reconstructions) are unnecessary to explain the experimental observations \cite{Thiel06}.
If the exposed AlO$_2$ surface is  passivated, our model predicts that electron transfer is suppressed and a 2DEG can in principle be observed, even for LAO layers below the critical thickness.  This explains why some experiments have observed conducting interfaces for LAO thickness of less than 4 unit cells \cite{Huijben06}.

We now turn to the observed variation in 2DEG density with oxygen partial pressure during growth or annealing \cite{Herranz07,Basletic08}.
LAO layers are typically AlO$_2$-terminated and exhibit partially filled oxygen dangling bonds.  Given the high electronegativity of oxygen and the position of such dangling-bond states close to the VBM of LAO, a strong driving force exists to fill these surface states with electrons.  Modifications of the surface that remove oxygen dangling bonds suppress electron transfer to the surface, and hence lead to a higher 2DEG density.
Growth or annealing in an environment with low O$_2$ partial pressure results in surface reconstructions containing oxygen vacancies (or equivalently cation adatoms), effectively removing oxygen dangling-bond states from the surface.  The situation is then closer to the scenario of the passivated surface described in Fig.~\ref{fig:SP}(d), which shows that a high-density 2DEG can develop.  While such oxygen treatments cannot be expected to lead to full passivation, the predicted trend of lower oxygen pressure resulting in higher 2DEG density is definitely consistent with experiment, without having to invoke modification of or point-defect formation at the buried STO/LAO interface.  Indeed, we consider the latter unlikely due to the high formation energy of oxygen vacancies in $n$-type STO \cite{Sukit11}.

Conversely, annealing under high O$_2$ partial pressure leads to a higher density of oxygen-related surface states that will consume electrons from the 2DEG.  This explains the seemingly puzzling fact that attempts to perfect the structural quality of STO/LAO heterostructure by growing or annealing under high O$_2$ partial pressure often lead to high sheet resistance or insulating behavior at the interface \cite{Mannhart08,Siemons07,Kalabukhov07}.

Manipulating oxygen partial pressure may not be the most effective means of passivating the LAO surface.  Hydrogen tends to be a good passivating agent.  Indeed, first-principles calculations \cite{Son10} found that hydrogenation of the LAO surface leads to an increase in the 2DEG density, although we disagree with Son {\it et al.}'s interpretation \cite{Son10} that hydrogen donates electrons to the interface.  The correct picture, as argued above, is that hydrogen passivates the surface and in the process prevents electrons being drained away from the interface.  Experimental efforts to identify the most effective means of controlling and passivating LAO surfaces could be highly fruitful.

Another approach to prevent electron transfer to the surface is to provide a source of electrons to the surface, for instance by depositing a metal on top of the LAO.  The effect will depend on the metal used, specifically, on the work function of the metal relative to the electron affinity of STO.  Metals with work function larger than the electron affinity of STO (e.g., Au) will still result in suppressed 2DEG densities since they will not succeed in suppressing electrons draining away from the interface.  Metals with work functions equal to or smaller than the electron affinity of STO (such as Ti or Al) are needed to increase 2DEG densities.  These qualitative insights are confirmed by explicit SP simulations.  These effects have also been observed in recent first-principles calculations \cite{Pentcheva12},
although again the interpretation should not be that electrons are being transferred from the metal to the interface.
Ultimately, though, metal capping layers on STO/LAO heterostructures may be more of academic interest, since it may prevent experimental probing of the 2DEG and also prove incompatible with device applications.  Emphasis on surface passivation techniques, as described above, is a more promising route.

Finally, we note that our model of the fundamental physics at STO/LAO interfaces transcends the specific materials system being discussed here and is generally applicable to oxide interface.  For instance, it explains why a 2DEG with the full nominal density of 3.3$\times$10$^{-14}$ cm$^{-2}$ has been observed at SrTiO$_3$/GdTiO$_3$ interfaces \cite{Moetakef11}. GdTiO$_3$ (GTO) is composed of alternating positively charged (GdO)$^+$ planes and negatively charged TiO$_2^-$ planes, similar to LaAlO$_3$ (note that Ti has valence 3 in GTO).  An STO/GTO interface with TiO$_2$-GdO interfacial planes will therefore also act as a sheet of donors.  The difference with the STO/LAO case lies in the fact that no electric field occurs within the GTO layers.  Indeed, STO can be grown with high quality on top of GTO, and since the interfaces are always between TiO$_2$ and GdO planes the top and bottom interfaces of each GTO layer are identical by construction.  This symmetry prevents an electrostatic potential buildup, as shown in Figs.~\ref{fig:STOLAO} and \ref{fig:SP}(a) and allows the full 3.3$\times$10$^{-14}$ cm$^{-2}$ density to be present in the 2DEG.

Even in the absence of an STO overlayer, a buildup of electrostatic potential is unlikely in the GTO layer.  If GdO-terminated, both the interface and surface would exhibit donor-like behavior. If TiO$_2$-terminated, the electrons that are needed to fill acceptor states do not flow into deep-lying oxygen-derived states, but rather into a Ti-derived lower Hubbard band which lies not far below the CBM of GTO \cite{Moetakef11}.  This position of the Fermi level at the surface again suppresses a buildup of potential and maintains the full 2DEG density at the STO/GTO interface.

In summary, based on first-principles calculations and Schr{\"o}dinger-Poisson simulations we demonstrate that electronic conductivity at
the STO/LAO interface arises from electrons that are intrinsic to the interface.  This precludes the need to invoke other sources of electrons such as the top LAO surface (according to the polar catastrophe model \cite{Nakagawa06,Demkov08}), or oxygen vacancies acting as donors \cite{Siemons07}.
The suppression of the 2DEG density at STO/LAO interfaces has often been attributed to interfacial reconstructions (either atomic or purely electronic, based on mixed valence of Ti), which in turn were invoked as a consequence of a ``polar catastrophe''.  Our present results show there is no need for invoking such mechanisms.
They emphasize the need for measures to prevent electrons draining away form the interface, which can be accomplished by preventing an electrostatic potential buildup in the LAO layer.
Proposed strategies include passivation of the surface, or depositing metals with suitably low work functions.
These insights into the origin of carriers at the STO/LAO interface will pave the way to enhanced control of the 2DEG at the interface of complex oxides.

AJ and CVdW were supported by the U. S. Army Research Office (W911-NF-11-1-0232).  LB was supported by the
NSF MRSEC Program (DMR-1121053).
Computational resources were provided by the Center for Scientific Computing at the CNSI and MRL (an NSF MRSEC, DMR-1121053) (NSF CNS-0960316), and by the Extreme Science and Engineering Discovery Environment (XSEDE), supported by NSF (OCI-1053575 and DMR07-0072N).
We are grateful to G. Sawatzky, S. Stemmer, and P. Moetakef for fruitful discussions.


\begin{thebibliography}{20}
\expandafter\ifx\csname natexlab\endcsname\relax\def\natexlab#1{#1}\fi
\expandafter\ifx\csname bibnamefont\endcsname\relax
 \def\bibnamefont#1{#1}\fi
\expandafter\ifx\csname bibfnamefont\endcsname\relax
 \def\bibfnamefont#1{#1}\fi
\expandafter\ifx\csname citenamefont\endcsname\relax
 \def\citenamefont#1{#1}\fi
\expandafter\ifx\csname url\endcsname\relax
 \def\url#1{\texttt{#1}}\fi
\expandafter\ifx\csname urlprefix\endcsname\relax\def\urlprefix{URL }\fi
\providecommand{\bibinfo}[2]{#2}
\providecommand{\eprint}[2][]{\url{#2}}

\bibitem{Ohtomo04}
A. Ohtomo, H. Y. Hwang,
Nature {\bf 427}, 423 (2004).

\bibitem{Mannhart08}
J. Mannhart,  D. H. A. Blank, H. Y. Hwang, A. J. Millis, J. M. Triscone,
MRS Bull. {\bf 33}, 1027 (2008).

\bibitem{Huijben09}
A. Huijben, G. Brinkman, G. Koster, G. Rijnders, H. Hilgenkamp, D. H. A. Blank,
Adv. Mater. {\bf 21}, 1665 (2009).


\bibitem{Brinkman07}
A. Brinkman, M. Huijben, M. van Zalk, U. Zeitler, J. C. Maan, W. G. van der Wiel, G. Rijnders, D. H. A. Blank, H. Hilgenkamp,
Nature Mater. {\bf 6}, 493 (2007).

\bibitem{Reyren07}
N. Reyren, S. Thiel, A.D. Caviglia, L. Fitting Kourkoutis, G. Hammerl, C. Ricther, C.W. Schneider, T. Kopp, A. S. Ruetschi, D. Jaccard, M. Gabay, D.A. Muller, J. M. Triscone, J. Mannhart,
Science {\bf 317}, 1196 (2007).

\bibitem{Li11}
L. Li, C. Richter, J. Mannhart, R. C. Ashoori,
Nature Phys. {\bf 7}, 762 (2011).

\bibitem{Hwang12}
H. Y. Hwang, Y. Iwasa, M. Kawasaki, B. Keimer, N. Nagaosa, Y. Tokura,
Nature Mater. {\bf 11}, 103 (2012).

\bibitem{Park06}
M. S. Park, S. H. Rhim, A. J. Freeman,
Phys. Rev. B {\bf 74}, 205416 (2006).

\bibitem{Siemons07}
W. Siemons, G. Koster, H. Yamamoto, W. A. Harrison, G. Lucovsky, T. H. Geballe, D. H. A. Blank, M. R. Beasley,
Phys. Rev. Lett. {\bf 98}, 196802 (2007).

\bibitem{Demkov08}
J. Lee, and A. A.  Demkov,
Phys. Rev. B {\bf 78}, 193104 (2008).

\bibitem{Popovic08}
Z. S. Popovi{\'c}, S. Satpathy, and R. M. Martin,
Phys.Rev.Lett. {\bf 101}, 256801 (2008).

\bibitem{Pentcheva10}
R. Pentcheva, W. E. Pickett,
J. Phys.: Condens. Matter {\bf 22}, 043001 (2010).


\bibitem{Bristowe11}
N. C. Bristowe, P. B. Littlewood, E. Artacho,
Phys. Rev. B {\bf 83}, 205405 (2011).

\bibitem{Sukit11}
Y. Li, S. N.  Phattalung, S. Limpijumnong, J. Kim, J. Yu,
Phys. Rev. B {\bf 84}, 245307 (2011) .


\bibitem{Kalabukhov07}
A. Kalabukhov, R. Gunnarsson, J. B{\"o}rjesson, E. Olsson, T, Claeson, D. Winkler,
Phys. Rev. B {\bf 75}, 121404(R) (2007).

\bibitem{Basletic08}
M. Basletic, J.-L. Mauric, C. Carr{\'e}t{\'e}ro, G. Herranz, O. Copie, M. Bibes, E. Jacquet, K., Bouzehouane, S. Fusil, A. Barth{\'e}l{\'e}my,
Nature Mater. {\bf 7}, 621 (2008).

\bibitem{Thiel06}
S. Thiel, G. Hammerl, A. Schmehl, C. W. Schneider, J. Mannhart,
Science {\bf 313}, 1942 (2006).

\bibitem{Herranz07}
G. Herranz, M. Basletic, M. Bibes, C. Carr{\'e}t{\'e}ro, E. Tafra, E. Jacquet, K. Bouzehouane,C. Deranlot, A. Hamzic, J. M. Broto, A. Barth{\'e}l{\'e}my, A. Fert,
Phys. Rev. Lett. {\bf 98}, 216803 (2007).

\bibitem{Huijben06}
M. Huijben, G. Rijnders, D. H. A. Blank, S. Bals, S. Van Aert, J. Verbeeck, G. Van Tendeloo, A. Brinkman, H. Hilgenkamp,
Nature Mater. {\bf  5}, 556 (2006).

\bibitem{Nakagawa06}
N. Nakagawa, H. Y. Hwang, D. A. Muller,
Nature Mater. {\bf 5}, 204 (2006).


\bibitem{Qiao10}
L. Qiao, T. C. Droubay, V. Shutthanandan, Z. Zhu, P. V. Sushko, S. A. Chambers,
Phys.: Condens. Matter {\bf 22}, 312201 (2010).


\bibitem{dft}
P. Hohenberg, W. Kohn,
Phys. Rev. {\bf 136}, B864 (1964);
W. Kohn, L. J. Sham,
Phys. Rev. {\bf 140}, A1133 (1965).

\bibitem{kressepaw}
G. Kresse, D. Joubert,
Phys. Rev. B {\bf 59}, 1758 (1999).


\bibitem{vasp2}
G. Kresse, J. Furthm$\ddot{\rm u}$ller, J.
Comput. Mat. Sci. {\bf 6}, 15 (1996).

\bibitem{hse}
J. Heyd, G. E. Scuseria, M. Ernzerhof,
J. Chem. Phys. {\bf 118}, 8207 (2003); erratum: \textit{J. Chem. Phys.} {\bf 124}, 219906 (2006).


\bibitem{varley12}
J. B. Varley, A. Janotti, C. Franchini, and C. G. Van de Walle,
Phys. Rev. B {\bf 85}, 081109 (2012).

\bibitem{nextnano}
A. Trellakis, T. Zibold, T. Andlauer, S. Birner, R. K. Smith, R. Morschl, P. J. Vogl,
Comput. Electron. \textbf{5}, 285 (2006); www.nextnano.de/nextnano3.

\bibitem{Son10}
W. Son, E. Cho, J. Lee, S. Han,
J. Phys. Cond. Matt. {\bf 22}, 315501 (2010).

\bibitem{Pentcheva12}
R. Arras, V. G. Ruiz, W. E. Pickett, R. Pentcheva,
Phys. Rev. B {\bf 85}, 125404 (2012).

\bibitem{Moetakef11}
P. Moetakef, T. A. Cain, D. G. Ouellette, J. Y. Zhang, D. O. Klenov, A. Janotti, C. G. Van de Walle, S. Rajan, S. J. Allen, S. Stemmer,
Appl. Phys. Lett. {\bf 99}, 232116 (2011).



\end{thebibliography}

\end{document}